\documentclass[aps,preprint]{revtex4}

\usepackage[dvips]{graphicx}
\usepackage[english]{babel}
\usepackage[T1]{fontenc}
\usepackage{mathrsfs}

\usepackage{amsxtra}
\usepackage{amsopn}
\usepackage{latexsym}
\usepackage[mathcal]{eucal}
\usepackage{mathtools}

\newcommand{\BE}{\begin{equation}}
\newcommand{\EE}{\end{equation}}
\newcommand{\BA}{\begin{align}}
\newcommand{\EA}{\end{align}}
\newcommand{\Tr}{\mathrm Tr}
\newcommand{\nn}{\nonumber}

\newcommand{\ppp}{ \frac{{\rm d}^4p}{(2\pi)^4}}

\newcommand{\fsl}[1]{\ensuremath{\mathrlap{\!\not{\phantom{#1}}}#1}}

\begin{document}

\title{From condensed matter to QCD: a journey through gauge\\ 
theories on board of a variational tool}
\author{Fabio Siringo}

\affiliation{Dipartimento di Fisica e Astronomia 
dell'Universit\`a di Catania,\\ 
INFN Sezione di Catania,
Via S.Sofia 64, I-95123 Catania, Italy}

\begin{abstract}
Starting with a review of the thermal fluctuations in superconductors,
the Gaussian Effective Potential is shown to be a powerful variational tool for
the study of the breaking of symmetry in gauge theories. A novel re-derivation
of the massive expansion for QCD is presented, showing its variational nature and its
origin from the Gaussian potential that also provides a variational proof for
chiral symmetry breaking and dynamical generation of a gluon mass.\\

Chapter 24 in {\it Correlations in Condensed Matter under Extreme 
Conditions: A tribute to Renato Pucci on the occasion of his 70th 
birthday}, edited by G. G. N. Angilella and A. La Magna (New York, 
Springer, 2017).
\end{abstract}





\maketitle

\section{Introduction}\label{sec1}

Since 1873, when Lord Rayleigh\cite{rayleigh} described a variational method for calculating the
frequencies of mechanical systems, the Rayleigh-Ritz method has
become an important tool for the approximate solution of physical problems in
quantum mechanics and quantum field theory. My personal experience on variational
methods dates back to 1985, when I was a graduate student of Professor Renato Pucci's.
He proposed to put an hydrogen molecule inside a rigid box and evaluate the
energy. His key idea was the insertion of a dielectric constant for simulating the
effects of the other molecules as if it were in a very dense phase under high pressure\cite{pucci}. 
That very physical idea was successful and since then Renato Pucci has been contributing to
the physics of solids under pressure with many  model calculations based on remarkable physical ideas.
Hydrogen was believed to become a superconductor in its solid phase under pressure and the fascinating
Anderson-Higgs mechanism of gauge symmetry breaking was one of the milestones in Professor Pucci's teaching.
That is where my personal journey has begun, going from the scalar $U(1)$ gauge 
theory of superconductivity\cite{gaussian,interpolation},
through the $SU(2)\times U(1)$ theory of weak interactions\cite{su2,LR,var,light,bubble}, up to $SU(3)$ 
theory and QCD\cite{sigma,sigma2,gep2,varqed,varqcd,genself,ptqcd2}. Still collaborating with
Renato Pucci in 2003, we found that a variational tool like the Gaussian Effective Potential (GEP) can describe
the thermal fluctuations of a superconductor in its broken-symmetry phase\cite{gaussian,interpolation}. 
While the same variational tool had
been very successful for describing the breaking of symmetry in a scalar theory\cite{stevenson}, its potentiality
in the study of gauge theories were not fully explored yet. The idea was then developed through several 
papers attempting to enlarge the gauge group\cite{su2,LR}, introduce fermions\cite{AF,HT} 
and eventually describe other mechanisms of
symmetry breaking, like the chiral symmetry breaking of 
QCD\cite{ptqcd2,ptqcd0,ptqcd,analyt,scaling}
where the gluon and quark masses emerge without any breaking of the gauge symmetry.

In this contribution, after reviewing the use of the GEP for the study of superconductivity\cite{gaussian,interpolation},  
the {\it massive expansion}\cite{ptqcd,ptqcd2,analyt} is re-derived from the GEP and shown to
be a powerful variational tool for addressing the problem of mass generation in Yang-Mills theories and
QCD, even when the gauge symmetry is not broken. While the massive expansion provides an analytical description of
the propagators of QCD from first principles\cite{ptqcd,ptqcd2} 
and is in remarkable agreement with the data of lattice simulations\cite{analyt,scaling}, its 
variational nature is hided and disguised to look like a perturbative method. The present novel and alternative
derivation of the massive expansion illustrate its direct origin from the GEP. Moreover, in chiral QCD the GEP 
provides a variational proof of chiral symmetry breaking and dynamical generation of the gluon mass.

\section{GEP and Superconductivity}
\label{sec2} 

The standard 
Ginzburg-Landau (GL) effective Lagrangian still provides the best 
framework for a general description of the phenomenology of $U(1)$
gauge symmetry breaking and superconductivity (the Anderson-Higgs mechanism).
The GL action can be seen as a power expansion of
the exact action around the critical point and is
recovered by any microscopic theory around the transition.
The Gaussian fluctuations can be studied by the GEP for $U(1)$ scalar electrodynamics
in three space dimensions\cite{gaussian,interpolation} where it represents the standard static
GL effective model of superconductivity. 
 
At variance with the approach of 
Ibanez-Meier et al.\cite{ibanez} who computed the GEP by use of a general
covariant gauge, we work in unitarity gauge, in order to make the physical content of the theory
more evident.
It has been shown to be formally equivalent to a full
gauge-invariant method once all the gauge degrees of freedom have
been integrated out\cite{mansfield}. The variational method
provides a way to evaluate both the correlation length $\xi$ and the
penetration depth $\lambda$ as a solution of coupled equations.
The GL parameter $\kappa_{GL}=\lambda/\xi$ is found to be temperature
dependent in contrast to the simple mean-field description and its
behaviour turns out to be in perfect agreement with many
experimental data\cite{gaussian,interpolation}.

Let us consider the standard static GL action\cite{kleinert}
\BE
S=\int d^{3}x \left[\frac{1}{4} F_{\mu \nu}F^{\mu
\nu}+\frac{1}{2}(D_{\mu}\phi)^{*}(D^{\mu}\phi)+
\frac{1}{2}m^{2}_{B}\phi^{*}\phi+
\lambda_{B}(\phi^{*}\phi)^{2}\right].
\label{gl}
\EE
where $\phi$ is a complex (charged) scalar field, its covariant
derivative is defined according to
\BE
D_{\mu}\phi = \partial_{\mu}+ie_{B} A_{\mu}
\label{derivative}
\EE
and $\mu,\nu=1,2,3$ run over the three space dimensions.
The magnetic field components are
$F_{\mu\nu}=\partial_\mu A_\nu-\partial_\nu A_\mu$. 
We may assume a transverse gauge $\nabla\cdot{\bf A}=0$, and then
switch to unitarity gauge in order to make $\phi$ real.

By a pure variational argument\cite{interpolation}
the longitudinal gauge field can be integrated out 
yielding the effective action
\BE
S=\int d^{3} x \left[
\frac{1}{2}(\nabla\phi)^{2}+\frac{m_{B}^{2}}{2}\phi^{2}
+\lambda_{B}\phi^{4}+\frac{e_{B}^{2}\phi^{2} A^2}{2}+
\frac{1}{2}({\nabla} \times {\bf A})^{2} 
+\frac{(\nabla\cdot{\bf A})^{2}}{2\epsilon}
\right].
\label{su2}
\EE
The partition function is expressed as a functional integral
over the real scalar field $\phi$ and the generic three-dimensional
vector field ${\bf A}$, with the extra prescription that the
parameter $\epsilon$ is set to zero at the end of the
calculation. As usual, the free energy (effective potential) follows by
inserting a source term and by a Legendre transformation\cite{gaussian,interpolation}.

The GEP may be evaluated
by the $\delta$ expansion method\cite{ibanez,stancu} and
is a variational
estimate of the exact free energy.
We introduce a shifted field
\BE
\tilde{\phi}=\phi-\varphi
\EE
then we split the Lagrangian into two parts
\BE
{\cal L}={\cal L}_0+{\cal L}_{int}
\EE
where ${\cal L}_0$ is the sum of two free-field terms
describing a vector field $A_\mu$ with mass $\Delta$ and
a real scalar field $\tilde \phi$ with mass $\Omega$:
\BE
{\cal L}_{0}=\left[
+\frac{1}{2}({\nabla} \times {\bf A})^{2} 
+\frac{1}{2} \Delta^{2} A_\mu A^\mu
+\frac{({\nabla}\cdot{\bf A})^{2}}{2\epsilon}
\right]
+\left[
\frac{1}{2}({\nabla}\tilde{\phi})^{2}
+\frac{1}{2}\Omega^{2}\tilde{\phi}^{2}
\right].
\label{L0}
\EE
The interaction then reads
\begin{align}
{\cal L}_{int}&=
v_{0}+v_{1}\tilde{\phi}+v_{2}\tilde{\phi}^{2}+v_{3}\tilde{\phi}^{3}
+v_{4}\tilde{\phi}^{4}+\nonumber\\
&+\frac{1}{2}\left(e_{B}^{2}\varphi^{2}-\Delta^{2}\right)A_\mu A^\mu
+e_{B}^{2}\varphi A_\mu A^\mu\tilde{\phi}
+\frac{1}{2}e^{2}_{B}A_\mu A^\mu\tilde{\phi}^{2}
\end{align}
where
\begin{align}
v_{0} &= \frac{1}{2}m_{B}^{2}\varphi^{2}+\lambda_{B}\varphi^{4},\quad
v_{1} = m_{B}^{2}\varphi+4\lambda_{B}\varphi^{3},\nonumber\\
v_{2} &= \frac{1}{2}m_{B}^{2}+6\lambda_{B}\varphi^{2}-\frac{1}{2}\Omega^{2},\quad
v_{3} = 4\lambda_{B}\varphi,\quad
v_{4} = \lambda_{B}.
\end{align}

The non conventional splitting of the Lagrangian has two important effects: arbitrary mass parameters
are inserted in the {\it free} part; mass counterterms are inserted in the interaction in order
to leave the Lagrangian unmodified. Then the standard perturbation theory is used for determining the
first-order effective potential. The sum of vacuum graphs up to first order yields the free-energy
density
\begin{align}
V_{eff}[\varphi] &= I_{1}(\Omega)+2I_{1}(\Delta)+\nn\\
&+ \left[
\lambda_{B}\varphi^{4}+\frac{1}{2}m_{B}^{2}\varphi^{2}+\frac{1}{2}
\left\{
m^{2}_{B}-\Omega^{2}+12\lambda_{B}\varphi^{2}+6\lambda_{B}I_{0}(\Omega)
\right\} I_{0}(\Omega) \right] \nn\\
&+\left( e_B^{2}\varphi^{2}+e_B^{2}I_{0}(\Omega)-\Delta^{2} \right)
I_{0}(\Delta)
\label{veff}
\end{align}
where the divergent integrals $I_n$ are defined according to
\BE
I_0 (M)=\int\frac{d^3 k}{(2\pi)^3} \frac{1}{M^2+k^2}, \quad
I_1 (M)=\frac{1}{2}\int\frac{d^3 k}{(2\pi)^3} \ln(M^2+k^2)
\label{In}
\EE
and must be regularized somehow.

The free energy (\ref{veff}) now depends on the mass parameters
$\Omega$ and $\Delta$. Since none of them was present in the original
GL action of Eq.(\ref{su2}), the free energy should not depend on them,
and the {\it minimum sensitivity} method\cite{minimal} 
can be adopted in order to fix the masses: the free energy is required
to be stationary for variations of $\Omega$ and $\Delta$. On the other
hand the stationary point can be shown to be a minimum for the free
energy and the method is equivalent 
to a pure variational method\cite{ibanez}.
At the stationary point the masses give the inverse correlation lengths
for the fields, the so called coherence length $\xi=1/\Omega$ and 
penetration depth $\lambda=1/\Delta$.

The stationary conditions
\BE
\frac{\partial V_{eff}}{\partial\Omega^{2}}=0, \qquad
\frac{\partial V_{eff}}{\partial\Delta^{2}}=0
\EE
give two coupled gap equations:
\BE
{\Omega}^{2}=12 \lambda_{B}
I_{0}({\Omega})+m^{2}_{B}+12 \lambda_{B}\varphi^{2}+2
e^{2}_{B}I_{0}({\Delta})
\label{gap1}
\EE
\BE
{\Delta}^{2}=e^{2}_{B}\varphi^{2}+e^{2}_{B}I_{0}({\Omega}).
\label{gap2}
\EE
For any value of $\varphi$, the equations must be solved 
numerically and the minimum-point values $\Omega$
and $\Delta$ must be inserted back into Eq.(\ref{veff})
in order to get the Gaussian free energy 
$V_{eff}(\varphi)$
as a function of the order parameter 
$\varphi$.
For a negative and
small enough $m_B^2$, we find that $V_{eff}$ has a minimum at a 
non zero value of $\varphi=\varphi_{min}>0$, thus indicating that the
system is in the broken-symmetry superconducting phase.
Of course the masses $\Omega$, $\Delta$ only take their physical
value at the minimum of the free energy $\varphi_{min}$.
That point may be found by requiring that
\BE
\frac{\partial V_{eff}}{\partial\varphi^{2}}=0
\label{phimin}
\EE
where as usual the partial derivative is allowed as far as the
gap equations (\ref{gap1}),(\ref{gap2}) are satisfied\cite{stevenson}.
The condition (\ref{phimin}) combined with the gap equation (\ref{gap1})
yields the very simple result
\BE
\varphi_{min}^{2}=\frac{{\Omega^{2}}}{8\lambda_{B}}.
\label{phimin2}
\EE
However, we notice that here the mass $\Omega$ must be found
by solution of the coupled gap equations. Thus Eqs.(\ref{phimin2}),
(\ref{gap1}) and (\ref{gap2}) must be regarded as a set of
coupled equations and must be solved together in order to find
the physical values for the correlation lengths and the order
parameter.

Insertion of Eq.(\ref{phimin2}) into Eq.(\ref{gap2}) yields a
simple relation for the GL parameter $\kappa_{GL}$
\BE
\kappa_{GL}^{2} = \left( \frac{\lambda}{\xi}
\right)^{2}=\kappa_0 \>\frac{1}
{\displaystyle 1+\frac{I_{0}(\Omega)}{\varphi_{min}^{2}} }
\label{kappa}
\EE
where $\kappa_0=e_B^2/(8\lambda_B)$ is the mean-field GL parameter
which does not depend on temperature. Eq.(\ref{kappa}) shows that
the GL parameter is predicted to be temperature dependent
through the non trivial dependence of $\Omega$ and $\varphi_{min}$.
At low temperature, where the order parameter $\varphi_{min}$ is
large, the deviation from the mean-field value $\kappa_0$ is
negligible. Conversely, close to the critical point, where the
order parameter is vanishing, the correction factor in Eq.(\ref{kappa})
becomes very important\cite{gaussian,interpolation}.

\begin{figure}[b] 
\centering
\includegraphics[width=0.35\textwidth,angle=-90]{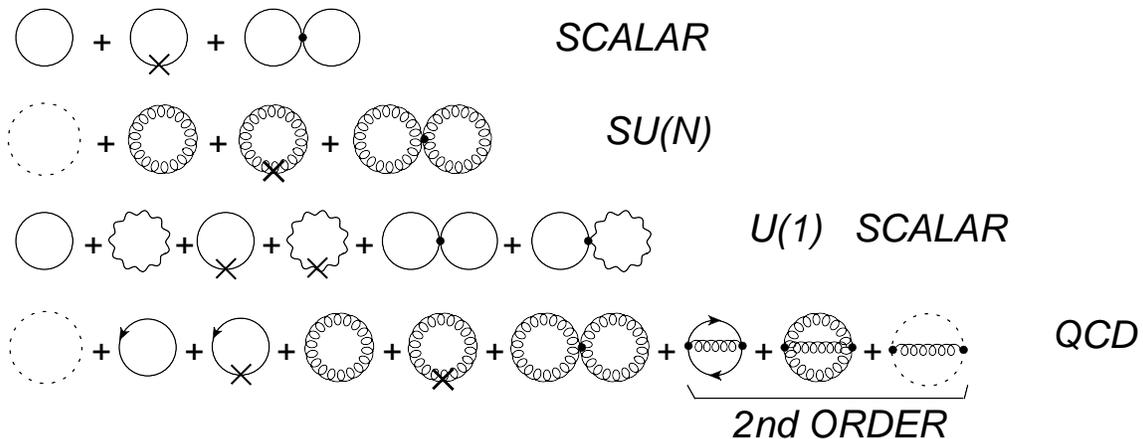}
\caption{Vacuum graphs contributing to the GEP for different theories}
\label{fig1}
\end{figure}

It is instructive to look at the effective potential in the limit $\varphi\to 0$
of the unbroken-symmetry phase and at the {\it chiral} point $m_B=0$ where the original
Lagrangian is scaleless. In that limit Eq.(\ref{veff}) reads
\begin{align}
V_{eff}[0]&= \left[ I_{1}(\Omega)+2I_{1}(\Delta)\right]
-\frac{1}{2}\left[\Omega^{2}I_{0}(\Omega)+2\Delta^{2} I_{0}(\Delta)\right]\nn\\
&+3\lambda_{B}\left[I_{0}(\Omega)\right]^2   
+e_B^{2}I_{0}(\Omega)I_{0}(\Delta)
\label{v0}
\end{align}
and is a function of the mass parameters. Its minimum might fall at a finite set of masses $\Delta_0$, $\Omega_0$
yielding a generation of mass from a scaleless Lagrangian. That property turns out to be useful for
addressing the problem of mass generation in chiral QCD where the gauge symmetry is not broken.
Moreover, we observe that all the terms in Eq.(\ref{v0}) arise from the sum of the vacuum graphs up to
first order, as shown in Fig.~\ref{fig1}, where the internal lines are the massive propagators that can 
be read from the free-particle Lagrangian ${\cal L}_0$ of Eq.(\ref{L0}). We obtain a {\it massive} expansion,
with massive free-particle propagators in the loops, from a massless Lagrangian.

\section{The Gaussian Effective Potential revisited}\label{sec3}

The massive expansion can be seen as an expansion around the
vacuum of massive particles. The search for the best vacuum is the aim of  the
GEP that has been studied by several authors, mainly in the context of spontaneous
symmetry breaking and scalar theories. While the GEP is a genuine variational method, several extensions to
higher orders have been proposed. However, being a first-order approximation, the GEP fails to predict any useful
result for the fermions of the standard model, because of the minimal gauge interaction that requires a
second order graph at least\cite{HT,gep2}. Even the idea of an expansion around the optimized vacuum of the GEP 
is not new\cite{tedesco} but has not been developed further.

In the next section, the pure variational nature of the GEP is used as a tool 
for demonstrating that the standard vacuum of QCD is unstable towards the vacuum of massive gluons and quarks.
Expanding around the optimized vacuum we recover the massive expansion that has been recently developed for pure
Yang-Mills theory\cite{ptqcd0,ptqcd,ptqcd2}. Thus, the unconventional massive perturbative expansion can be seen 
to emerge from the GEP formalism in a natural way. 

One of the important merits of the GEP is its paradox of being a pure variational method disguised as a
perturbative calculation,  making use of the standard graphs of perturbation theory. In this section
we set the formalism of the expansion, starting with the simple scalar theory and then moving towards Yang-Mills
theory and QCD in the next section.

Let us revise briefly the method for the simple case of a self-interacting scalar theory\cite{stevenson}
where the effective potential is given by three vacuum graphs as shown in Fig.~\ref{fig1} (to be compared with the
six graphs of the scalar electrodynamics of Sec \ref{sec2}). 
The Lagrangian density reads
\BE
{\cal L}=\frac{1}{2} \phi \left(-\partial^2-m_B^2\right)\phi-\frac{\lambda}{4!}\phi^4
\label{Lphi}
\EE
where $m_B$ is a bare mass. We can then split the total Lagrangian as ${\cal L}={\cal L}_0+{\cal L}_{int}$
where the trial free part is
\BE
{\cal L}_0=\frac{1}{2} \phi \left(-\partial^2-m^2\right)\phi
\label{Lfree}
\EE
and describes a free scalar particle with a trial mass $m\not= m_B$.
The interaction follows as
\BE
{\cal L}_{int}=-\frac{\lambda}{4!}\phi^4-\frac{1}{2} \left(m_B^2-m^2\right)\phi^2
\label{Lint}
\EE
so that the total Lagrangian has not been changed.
If we neglect the interaction, then a free Hamiltonian ${\cal H}_0$ is derived from ${\cal L}_0$ and its
ground state $\vert m\rangle$ satisfies
\BE
{\cal H}_0 \> \vert m\rangle=E_0(m)\> \vert m\rangle
\EE
and depends on the trial mass $m$.
Restoring the interaction ${\cal L}_{int}$, the full Hamiltonian reads ${\cal H}={\cal H}_0+{\cal H}_{int}$ and
by standard perturbation theory, the first-order energy of the ground state reads
\BE
E_1 (m)= E_0(m) + \langle m\vert {\cal H}_{int} \vert m\rangle=\langle m\vert {\cal H} \vert m\rangle
\label{E1}
\EE
and is equivalent to the first-order effective potential $V_1 (m) $ evaluated by standard perturbation theory
with the interaction ${\cal L}_{int}$. Thus, the stationary condition
\BE
\frac{\partial V_1 (m)}{\partial m}=\frac{\partial E_1 (m)}{\partial m}=0
\label{stat}
\EE
gives the best value of $m$ that minimizes the vacuum energy of the ground state $\vert m\rangle$.
While being a pure variational method, the first-order effective potential $V_1(m)=E_1(m)$ can be evaluated 
by the sum of all the vacuum graphs up to first order (the three loop graphs in Fig.~\ref{fig1}). 
The resulting optimized effective potential is the GEP. Usually, the effective potential
is evaluated for any value of the average $\varphi=\langle \phi\rangle$ and the best $m$ also depends on that average.
If the symmetry is not broken, then the minimum of the effective potential is at $\varphi=0$ where
$V_1 (m)$ is a function of the trial mass, to be fixed by the stationary condition Eq.(\ref{stat}).
We assume that the gauge symmetry is not broken in QCD so that $V_1(m)$ at $\varphi=0$ is the effective potential 
we are interested in.
The variational nature of the method ensures that the true vacuum energy is smaller than the minimum of $V_1(m)$.
At the minimum, $\vert m\rangle$ provides an approximation for the vacuum and is given by the vacuum of a free
massive scalar particle with mass equal to the optimized mass parameter $m\not=m_B$.
Of course, the optimal state $\vert m\rangle$ is just a first approximation and the actual vacuum is 
much richer. However, we expect that a perturbative expansion around that approximate vacuum would be the
best choice for the Lagrangian ${\cal L}$, prompting towards an expansion with an interaction ${\cal L}_{int}$
and a free part ${\cal L}_0$ that depend on $m$ and can be optimized by a clever choice of the parameter $m$.
Different strategies have been proposed for the optimization, ranging from the stationary condition of the GEP,
Eq.(\ref{stat}), to Stevenson's principle of minimal sensitivity\cite{minimal}. 
A method based on the minimal variance has been recently proposed for QCD and 
other gauge theories\cite{sigma,sigma2,gep2,varqed,varqcd}. In all those approaches, the underlying idea is that
an optimal choice of $m$ could minimize the effect of higher orders in the expansion. Since the total Lagrangian
does not depend on $m$, the physical observables are expected to be stationary at the optimal $m$, thus suggesting
the use of stationary conditions for determining the free parameter. As a matter of fact, 
if all graphs were summed up exactly, then the dependence on $m$ would cancel in the final result, so that the 
strength of that dependence measures the weight of the neglected graphs at any order.

Leaving aside the problem of the best choice of $m$, we observe that at $\varphi=0$ the calculation of
the first-order effective potential $V_1(m)$ is quite straightforward and follows from the first-order expansion of
the effective action $\Gamma(\varphi)$
\BE
e^{i\Gamma(\varphi)}=\int_{1PI} {\cal D}_{\phi} e^{i S_0(\phi+\varphi)+i S_{int}(\phi+\varphi)} 
\EE
where the functional integral is the sum of all one-particle irreducible (1PI) graphs and $S=S_0+S_{int}$
is the action. The effective potential then follows as $V_1(m)=-\Gamma(0)/{\cal V}_4$ where ${\cal V}_4$ is
a total space-time volume. Moreover, being interested in the {\it chiral} limit, let set $m_B=0$ in the
interaction Eq.(\ref{Lint}) and study a massless scalar theory.

The vertices of the theory can be read from ${\cal L}_{int}$ in Eq.(\ref{Lint}) where we set $m_B=0$ 
and are used in Fig.1 in the vacuum graphs up to first order. 
The usual four-point vertex  $-i \lambda$ is accompanied by
the counterterm $i m^2$ that is denoted by a cross in the graphs. This counterterm must be regarded as part of
the interaction so that the expansion in not loopwise and we find one-loop and two-loop graphs summed together in
the first-order effective potential. That is where the non-perturbative nature of the method emerges since the
expansion in not in powers of $\lambda$ but of the whole interaction ${\cal L}_{int}$.
The zeroth order (massive) propagator $ i \Delta_m$ follows from ${\cal L}_0$ 
\BE
i \Delta_m (p)=\frac{i}{p^2-m^2}
\EE
and is shown as a straight line in the vacuum graphs.

The tree term is the classical potential and vanishes in the limit $\varphi\to 0$. The first one-loop graph
in Fig.1 gives the standard one-loop effective potential, containing some effects of quantum fluctuations.
It must be added to the second one-loop graph in Fig.1, the crossed graph containing one insertion of the
counterterm. It is instructive to see that the exact sum of all one-loop graphs with $n$ insertions of the
counterterm gives the standard vacuum energy of a massless particle. In other words, if we sum all the crossed
one-loop graphs the dependence on $m$ disappears and we are left with the standard one-loop effective potential
of Weinberg and Coleman\cite{WC} $V_{1L}^0=-\Gamma_{1L}^0/{\cal V}_4$ where $\Gamma_{1L}^0$ is the standard
one-loop effective action at $\varphi=0$
\BE
e^{i\Gamma_{1L}^0}=\int {\cal D}_{\phi} e^{i \int \frac{1}{2} \phi (-\partial^2)\phi \>{\rm d}^4x}
\sim\left[{\rm Det}(\Delta_0^{-1})\right]^{\displaystyle{-\frac{1}{2}}}
\label{G1L}
\EE
and $\Delta_0^{-1}=p^2$ is the free-particle propagator of a massless scalar particle.

Up to an additive constant, not depending on $m$, Eq.(\ref{G1L}) can be written as
\BE
V^0_{1L}=\frac{-i}{2{\cal V}_4}\Tr \log(\Delta_m^{-1}+m^2)
\EE
then expanding the log we obtain a massive expansion
\BE
V^0_{1L}=\frac{-i}{2{\cal V}_4} \Tr\left\{ \log(\Delta_m^{-1})+ m^2\Delta_m
-\frac{1}{2}m^2\Delta_m m^2\Delta_m+\cdots\right\}
\label{massexp}
\EE
that is shown pictorially in Fig.2 as a sum of crossed one-loop vacuum graphs. While the sum cannot depend on
$m$, if we truncate the expansion at any finite order we obtain a function of the mass parameter.
As a test of consistency, one can easily check that, once renormalized, the sum of all the crossed 
one-loop vacuum graphs in Fig.2 gives zero exactly.

\begin{figure}[b] 
\centering
\includegraphics[width=0.2\textwidth,angle=-90]{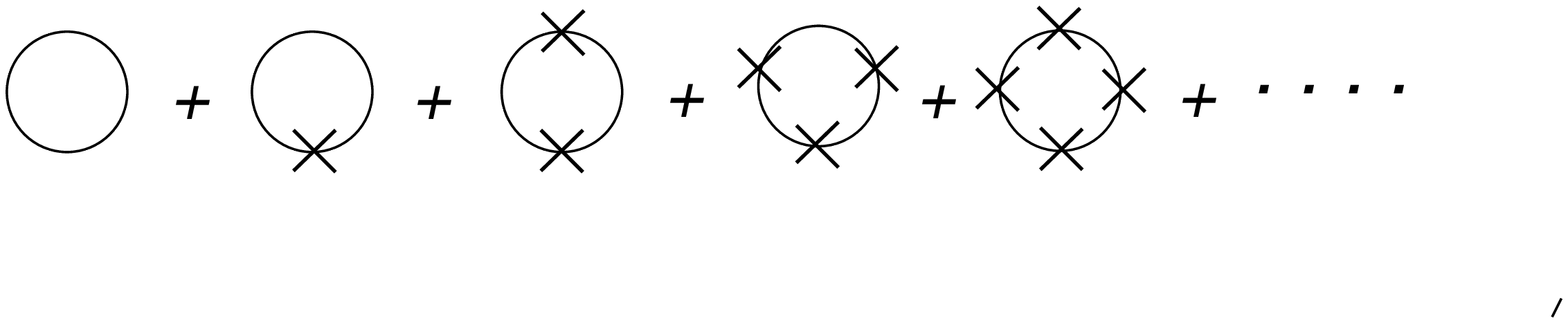}
\caption{Pictorial display of the right hand side of Eq.(\ref{massexp}).}
\label{fig2}
\end{figure}

The calculation of the GEP requires the sum of only the first two terms of Eq.(\ref{massexp}), 
the two one-loop graphs in Fig.1.  We cannot add
higher-order terms without spoiling the variational method since the average value of the Hamiltonian
in the trial state $\vert m\rangle$ is  $E_1(m)=V_1(m)$, according to Eq.(\ref{E1}).
Using the identity
\BE
\Delta_m=-\frac{\partial}{\partial m^2} \log(\Delta_m^{-1})
\EE
the sum of one-loop graphs in Fig.1 can be written as
\BE
V_{1L} (m)=\left( 1-\frac{\partial}{\partial m^2}\right) I_1(m)=I_1(m)-\frac{1}{2} m^2 I_0(m)
\label{V1L}
\EE
where the diverging integrals $I_1(m)$, $I_0 (m)$ generalize Eq.(\ref{In}) and are defined as
\BE
I_1(m)=\frac{1}{2i}\int \ppp \log(-p^2+m^2),\quad
I_0(m)=-i\int \ppp \frac{1}{-p^2+m^2}
\label{I}
\EE
so that 
\BE
\frac{\partial I_1 (m)}{\partial m^2}=\frac{1}{2} I_0(m).
\label{dI1}
\EE
We recognize $I_1 (m)$ as the standard one-loop effective potential of Weinberg and Coleman for a
massive scalar particle in the limit $\varphi\to 0$. This term contains the quantum fluctuations at
one-loop. The second term is a correction coming from the counterterm and arises because the exact
Lagrangian was massless. The calculation of the GEP also requires the two-loop graph in Fig.1 that
is first-order in $\lambda$. A lazy way to evaluate it is by substituting the vertex $im^2$ in
the crossed one-loop graph with the seagull one-loop self energy graph $-i \Sigma_{1L}$ that reads\cite{gep2}
\BE
\Sigma_{1L}=\frac{\lambda}{2} I_0 (m)
\EE
and adding a $1/2$ symmetry factor. The resulting two-loop term is
\BE
V_{2L} (m)=\frac{\lambda}{8} [I_0 (m)]^2.
\label{V2L}
\EE
The GEP follows as the sum $V_{1L}+V_{2L}$
\BE
V_{GEP}(m)=I_1(m)-\frac{1}{2}m^2 I_0(m) +\frac{\lambda}{8} [I_0 (m)]^2.
\label{GEP}
\EE
At this stage we just recovered the GEP in the limit $\varphi\to 0$ and Eq.(\ref{GEP}) agrees with
the well known GEP in that limit\cite{stevenson,stancu2,stancu,gep2} (also compare to Eq.(\ref{v0}) by
setting $4!\lambda_B=\lambda$, $\Omega=m$ and neglecting gauge field loops).

More precisely, $V_{GEP}$ is the GEP when $m$ is optimized by the stationary condition Eq.(\ref{stat})
that reads
\BE
\frac{\partial V_{GEP} (m)}{\partial m^2}=\frac{1}{2}\left(\frac{\partial I_0 (m)}{\partial m^2}\right)
\left[\frac{\lambda I_0(m)}{2}-m^2\right]=0
\label{factors}
\EE
yielding the usual gap equation of the GEP
\BE
m^2=\frac{\lambda I_0(m)}{2}.
\label{gap}
\EE
From a mere formal point of view, the GEP predicts the existence of a mass for the  massless scalar theory.
That is of special interest because for $m_B=0$ the Lagrangian in Eq.(\ref{Lphi}) has no energy scale, just
like Yang-Mills theory and QCD in the chiral limit. Thus, it can be regarded as a toy model for the more
general problem of mass generation and chiral symmetry breaking.

Actually, the integrals $I_0$, $I_1$ are badly diverging, and a mass scale arises from the regulator that must be
inserted in order to get a meaningful theory.  We can see that, in dimensional regularization, by
setting $d=4-\epsilon$, the integral $I_0$ is
\BE
I_0(m)=-\frac{m^2}{16\pi^2}\left[\frac{2}{\epsilon} + \log \frac{\bar\mu^2}{m^2}+1+{\cal O}(\epsilon)\right]
\EE
where $\bar \mu=(2\sqrt{\pi}\mu)\exp(-\gamma/2)$ is an arbitrary scale. Integrating Eq.(\ref{dI1})
and neglecting an integration constant (that does not depend on $m$)
\BE
I_1(m)=-\frac{m^4}{64\pi^2}\left[\frac{2}{\epsilon} + \log \frac{\bar\mu^2}{m^2}+\frac{3}{2}+{\cal O}(\epsilon)\right].
\EE
If we follow the usual approach of Weinberg and Coleman\cite{WC}, the divergences must be absorbed by the physical
renormalized parameters. Thus, let us define a {\it physical} renormalized energy scale $\Lambda$ as
\BE
\log \Lambda^2= \log \bar \mu^2+\frac{2}{\epsilon}+1
\label{Lambda}
\EE
and write the integrals $I_1$, $I_0$ as simply as
\begin{align}
I_0(m)&=\frac{m^2}{16\pi^2}\log \frac{m^2}{\Lambda^2}\nn\\
I_1(m)&=\frac{m^4}{64\pi^2}\left[\log \frac{m^2}{\Lambda^2}-\frac{1}{2}\right].
\label{IDR}
\end{align}
This approach is the same that is usually followed in lattice simulations of QCD: the lattice provides a
scale that can be changed without affecting the physical scale which remains fixed at a phenomenological value.
We assume that when $\epsilon\to 0$ the scale $\bar \mu$ also changes, keeping $\Lambda$ fixed at a physical
value which cannot be predicted by the theory, but must come from the phenomenology.

\begin{figure}[b] 
\centering
\includegraphics[width=0.55\textwidth,angle=-90]{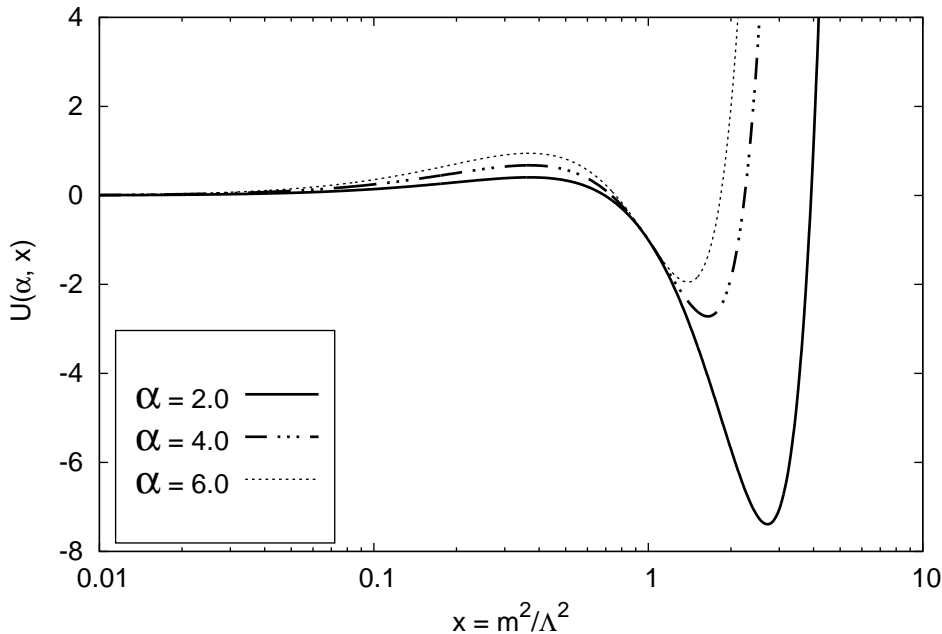}
\caption{The adimensional potential $U(\alpha,x)$ of Eq.(\ref{U}) is shown for different values
of the effective coupling $\alpha$.}  
\label{fig3}
\end{figure}

First of all, we observe that by our renormalization scheme the standard one-loop effective potential is recovered,
since that is equal to $I_1(m)$ in Eqs.(\ref{IDR}) and can be recognized as the mass-dependent term of the 
standard one-loop effective potential in the limit $\varphi\to 0$. That term has a relative maximum at $m=0$, 
is negative for $m<\Lambda\>\exp(1/4)$ and has the absolute minimum at $m=\Lambda$. Thus, the one-loop effective
potential would predict a massive vacuum if the symmetry were not broken and the physical vacuum were at $\varphi=0$.

The full renormalized GEP is finite in terms of the physical scale $\Lambda$ and can be written as
\BE
V_{GEP}(m)=\frac{\Lambda^4}{128\pi^2} U(\alpha, m^2/\Lambda^2)
\EE
where the adimensional potential $U(\alpha,x)$ is 
\BE
U(\alpha,x)=x^2\left[\alpha (\log x)^2-2\log x-1\right]
\label{U}
\EE
and $\alpha$ is the effective coupling $\alpha=\lambda/(16\pi^2)$.

The behavior of the potential $U(\alpha,x)$ is shown in Fig.3. For any coupling $\alpha$ the point
$x=0$ is a relative minimum while the potential has a relative maximum at $x=1/e$.
The absolute  minimum is at $x_0=\exp(2/\alpha)$ where $U(\alpha,x_0)=-x_0^2<0$. The two stationary points
$x=1/e$ and $x=x_0$ are the points where the first or second factor in Eq.(\ref{factors}) is zero, respectively.
Thus the absolute minimum $m^2/\Lambda^2=x_0=\exp(2/\alpha)$ is the solution of the gap equation, Eq.(\ref{gap}).
However, since the original theory has no scale, the quantitative value of $m$ remains arbitrary as it
depends on the unknown scale $\Lambda$. We can only predict that, since the GEP provides a genuine variational
approximation for the vacuum energy, the massless vacuum must be unstable towards the vacuum of a massive scalar 
particle with an exact effective potential $V_{exact} (m)\le V_{GEP}(m) <0$.

In the chiral limit, the GEP can be easily extended to more complex theories by just adding up the
graphs of Fig.\ref{fig1}. It is instructive to see how Eq.(\ref{v0}) can be recovered by the graphs for
scalar $U(1)$ electrodynamics. In the next section the GEP is evaluated for the chiral limit of QCD.

\section{QCD in the Chiral Limit}\label{sec4}

The full Lagrangian of QCD, including $N_f$ massless chiral Quarks, can be written as
\BE
{\cal L}_{QCD}={\cal L}_{YM}+\sum_{i=1}^{N_f}
\bar\Psi_i\left[i\fsl{\partial}-g{\ensuremath{\mathrlap{\>\not{\phantom{A}}}A_a}}\hat T_a\right]\Psi_i
\label{LQCD}
\EE
where ${\cal L}_{YM}$ is the full $SU(N)$ Yang-Mills Lagrangian, including a covariant gauge-fixing term
and the ghost terms arising from the Faddeev-Popov determinant.
The generators of $SU(N)$ satisfy the algebra
\BE
\left[ \hat T_a, \hat T_b\right]= i f_{abc} \hat T_c
\EE
with the structure constants normalized according to
\BE
f_{abc} f_{dbc}= N\delta_{ad}.
\label{ff}
\EE
In a background field $\tilde A_a^\mu$ the effective action $\Gamma (\tilde A)$ is the sum of 1PI graphs
that can be formally given by the functional integral
\BE
e^{i\Gamma( \tilde A)}=\int_{1PI} {\cal D}_{\Psi, A, \omega}\> e^{i S[A+\tilde A, \Psi, \omega]}
\EE
where $\omega_a$ are the ghost fields.
Assuming that the gauge symmetry is not broken, we are interested in the study of the limit $\tilde A\to 0$ and 
write the effective action as
\BE
e^{i\Gamma}=\int_{1PI} {\cal D}_{ A, \omega} e^{i S_{YM} [A, \omega] +i\Gamma_\Psi[A]}
\EE
where $S_{YM}$ is the action of pure Yang-Mills theory and the effective action $\Gamma_\Psi$ 
is given by a functional integral over quark fields
\BE
e^{i\Gamma_\Psi[A]}=\int {\cal D}_{ \Psi} e^{i \int \bar \Psi \>\hat D(A)\> \Psi \>{\rm d}^4x}
\EE
with the operator $\hat D(A)$ that is given by
\BE
\hat D(A)=i\fsl{\partial}-g{\ensuremath{\mathrlap{\>\not{\phantom{A}}}A_a}}\hat T_a.
\EE

The quark fields can be integrated exactly, yielding, up to a constant,
\BE
i\Gamma_\Psi(A)=\log {\rm Det}\hat D(A).
\label{GammaPsi}
\EE
While $S_{YM}$ contains the vertices of pure Yang-Mills theory, the expansion of $\Gamma_\Psi$ in
powers of $gA_a^\mu$ provides the standard insertions of quark-gluon vertices, yielding the usual
Feynman rules of QCD. Some vacuum graphs, up to second order and two loops, are shown in Fig.~\ref{fig1}.

As already noticed for the scalar theory, the calculation of the GEP requires the first-order effective potential
that results from the sum of connected vacuum graphs up to first-order. Thus we may focus on the one-loop
graphs in Fig.~\ref{fig1} and on the only first-order two-loop graph (the fourth for $SU(N)$ in Fig.~\ref{fig1}). 
All other graphs are
second order at least, starting from the other two-loop graphs of Fig.~\ref{fig1}.
Thus, at first order, the effective potential $V_1$ is just the sum of independent ghost, gluon and quark terms.
This is an important limit of the GEP that cannot take in due account the second-order graphs, leaving us with
a decoupled description of quarks, gluons and ghosts. 
We can write the first-order effective potential as
\BE
V_1=V_{YM}+V_\Psi
\EE
where the quark term contains only the one-loop zeroth-order vacuum graph that arises from Eq.(\ref{GammaPsi})
at $g=0$
\BE
V_\Psi=\frac{i}{{\cal V}_4} \log {\rm Det} \hat D_0
\label{VPsi}
\EE
having defined the zeroth-order operator $\hat D_0=i\fsl{\partial}$.
The Yang-Mills term $V_{YM}$ is the first-order effective potential of pure Yang-Mills theory 
and can be written as
\BE
V_{YM}=\frac{i}{{\cal V}_4} \log \int_{\rm 1st-order} {\cal D}_{ A, \omega} e^{i S_{YM} [A, \omega]}
\label{VYM}
\EE
and is given by the one-loop ghost graph plus the one-loop and two-loop gluon graphs in Fig.~\ref{fig1}.

At this stage, the whole calculation might seem to give trivial constant terms. 
However, we are interested in the {\it change} of these terms when a massive zeroth order propagator is
taken from the beginning for gluons and quarks. As already seen for the scalar theory, we have the freedom of
adding a mass term in the zeroth order Lagrangian provided that we subtract the same mass term in
the interaction. The resulting massive expansion contains new two-point vertices (the mass counterterms)
and their insertion in a graph does not change the number of loops but increases the order of the graph.
Moreover, the first-order vacuum graphs in Fig.~\ref{fig1} remain uncoupled when any number of counterterms is inserted,
so that we can study the change induced by the masses on $V_\Psi$ and $V_{YM}$ separately. It is
instructive to see how the massive expansion\cite{ptqcd,ptqcd2} of Yang-Mills theory emerges naturally 
in the calculation of the GEP and can be extended to chiral quarks.

\subsection{Pure Yang-Mills Theory}\label{sec41}

In a generic linear covariant $\xi$-gauge, the first-order effective potential $V_{YM}$ can be written as the sum
of the second and fourth graph in Fig.~\ref{fig1}, namely the zeroth order gluon loop and the first-order two-loop graph
which contains one insertion of the four-gluon vertex. We may drop the decoupled ghost loop that only gives an
additive constant to the effective potential.

By the same notation of Sec.\ref{sec3}, we denote by $V^0_{1L}$ the one-loop graph that gives the standard
one-loop effective potential in the limit of a vanishing background field
\BE
V^0_{1L}=\frac{i}{{\cal V}_4} \log \int {\cal D}_{ A} e^{i \int A_{a\mu} {\Delta^{-1}_0}^{\mu\nu}
A_{a\nu} {\rm d}^4x}
\EE
containing the quadratic part of $S_{YM}$ in Eq.(\ref{VYM}) written in terms of the gluon propagator
\BE
\Delta^{\mu\nu}_0 (p)=\Delta^T_0(p) t^{\mu\nu}(p)+\Delta^L_0 (p) l^{\mu\nu}(p)
\label{Delta}
\EE
where $t^{\mu\nu}$, $l^{\mu\nu}$ are the transversal and longitudinal Lorentz projectors
\BE
t^{\mu\nu}(p)=g^{\mu\nu}-\frac{p^\mu p^\nu}{p^2}, \qquad
l^{\mu\nu}(p)=\frac{p^\mu p^\nu}{p^2}
\label{proj}
\EE
and the corresponding free-particle scalar functions are
\BE
\Delta_0^T(p)=\frac{1}{-p^2},\qquad \Delta_0^L(p)=\frac{\xi}{-p^2}.
\EE
The determinant of $\Delta_0$ can then be written as a product of determinants in the
orthogonal Lorentz subspaces, ${\rm Det} \Delta_0=({\rm Det}\Delta_0^T) ({\rm Det}\Delta_0^L)$,
yielding
\BE
V^0_{1L}=\frac{i}{2{\cal V}_4}\left[\Tr \log \Delta_0^T+\Tr \log \Delta_0^L\right].
\EE

From now on, we work in the Landau gauge and take the limit $\xi\to 0$. In that limit
$\Delta^L_0\to 0$ and the longitudinal part gives an (infinite) additive constant that we drop.
The relevant part we will focus on reads
\BE
V^0_{1L}=\frac{N_A}{2 i}\int\ppp\log \left({\Delta_0^T}^{-1}\right)
\label{V01L}
\EE
where $N_A$ is a factor arising from the trace over color and Lorentz indices.

Following the same steps that lead to the GEP for a scalar theory, we may modify
the quadratic part of the Lagrangian, i.e. $\Delta_0^{-1}$ in $S_{YM}$, provided that we
add a counterterm to the total Lagrangian in order to leave it unchanged.
Thus we add a mass term to the transversal part $\Delta^T_0$, leaving the longitudinal 
part unmodified. That would be a reasonable choice in any gauge since the longitudinal
part $\Delta_0^L$ is left unmodified by the interaction at any order of perturbation theory.
We define a new massive zeroth-order propagator $\Delta_m^T$ as
\BE
{\Delta^T_m}^{-1}={\Delta^T_0}^{-1}+m^2=-p^2+m^2
\label{Deltam}
\EE
and insert the counterterm 
\BE
\delta {\cal L}_c= m^2 t^{\mu\nu} A_{a\mu} A_{a\nu}
\label{Lc}
\EE
in the Lagrangian density. Then we look at the change of the first-order effective potential
as a function of the mass parameter $m$, including the counterterm as a vertex of the theory.
The result is formally equivalent to that obtained for the massless scalar theory in Eq.(\ref{massexp})
and Fig.~\ref{fig2}. By insertion of the counterterm, the one-loop gluon loop gives rise to an infinite sum of
crossed loops where the straight line in Fig.~\ref{fig2} is now given by the massive propagator of Eq.(\ref{Deltam})
and the crosses denote the insertion of a two-point vertex $-i m^2 t^{\mu\nu}$. 
Even in a generic $\xi$-gauge the longitudinal part of the gluon propagator would not add any higher
order contribution because of the transversal projector in the counterterm. Since everything is transversal
in the Landau gauge, from now on we drop any projector $t^{\mu\nu}$ and the superscript $T$ in the 
transverse propagator. Writing $\log(\Delta_0^{-1})=\log (\Delta_m^{-1}-m^2)$ 
in Eq.(\ref{V01L}) and expanding the log,
the one-loop graph $V_{1L}^0$, that does not depend on $m$, reads
\BE
V^0_{1L}=\frac{N_A}{2 i}\int\ppp\left\{
\log \Delta_m^{-1} - \sum_{n=1}^{\infty}\frac{(m^2\Delta_m)^n}{n}\right\}.
\label{V01Lm}
\EE

As before, in order to evaluate the GEP we must truncate the expansion and retain terms up to
the first order, namely the zeroth-order gluon loop and the first order crossed loop that are the
first two graphs in Fig.~\ref{fig2}. Then, at first-order, the one loop effective potential is 
\BE
V_{1L} (m)=N_A\left( 1-\frac{\partial}{\partial m^2}\right) I_1(m)
\label{V1LA}
\EE
that is the same result of Eq.(\ref{V1L}) scaled by the trace factor $N_A$.

The GEP also includes the two-loop first-order graph, the fourth graph in Fig.~\ref{fig1} with the propagator
replaced by the massive propagator $\Delta_m$ and no insertions of the counterterm that would raise
the order of the graph. By the same argument that leads to Eq.(\ref{V2L}), the two-loop graph is
easily evaluated by substituting the vertex $-im^2$ in
the crossed one-loop graph with the one-loop seagull self energy graph $-i \Pi_{1L}$ that reads\cite{genself}
\BE
\Pi_{1L}=-\frac{9Ng^2}{4} I_0 (m)
\EE
and adding a $1/2$ symmetry factor. The resulting two-loop term is
\BE
V_{2L} (m)=\frac{9N_A N g^2}{16} [I_0 (m)]^2.
\label{V2LA}
\EE
Adding the one-loop terms the GEP reads
\BE
V_{GEP}(m)=N_A\left\{I_1(m)-\frac{1}{2}m^2 I_0(m) +\frac{9 N g^2}{16}[I_0 (m)]^2\right\}
\label{GEPA}
\EE
which is exactly the same result of Eq.(\ref{GEP}) for a scalar theory with 
an effective coupling $\lambda=9 N g^2/2$, scaled by the trace factor $N_A$.
Then by dimensional regularization, in the same scheme of Sec.~\ref{sec3}, the GEP 
of pure Yang-Mills theory can be written as
\BE
V_{GEP}(m)=\frac{\Lambda^4N_A}{128\pi^2} U(\alpha, m^2/\Lambda^2)
\label{GEPYM}
\EE
where the effective coupling $\alpha=\lambda/(16\pi^2)=9 N \alpha_s/(8\pi)$,
$\alpha_s=g^2/(4\pi)$ and $\Lambda$ is an unknown scale that must be fixed by
the phenomenology. The adimensional potential $U(\alpha, x)$ was defined in Eq.(\ref{U})
and shown in Fig.~\ref{fig3}.

\subsection{Including Chiral Fermions}\label{sec42}

The inclusion of a set of chiral quarks is straightforward. As shown in Fig.~\ref{fig1},
up to first order, the fermions are decoupled in the effective potential and we must just add the
two one-loop graphs for the quarks. Let us derive them by the same method of Sec.~\ref{sec41}.
For fermions, the standard one-loop effective potential $V_\Psi$ of Eq.(\ref{VPsi}) can be written
as
\BE
V_\Psi=\frac{i}{{\cal V}_4} \log {\rm Det} \left( \hat D_M+M\right)
\label{VPsiM}
\EE
where the {\it massive} inverse propagator $\hat D_M=\hat D_0-M$ and the parameter $M$ is an arbitrary trial
quark mass. The exact expansion of Fig.~\ref{fig2} is recovered again as
\BE
V_\Psi=\frac{i}{{\cal V}_4} \Tr\left[\log \hat D_M\right]
+\frac{i}{{\cal V}_4} \Tr\left[ 
\sum_{n=1}^{\infty}\frac{\left(\hat D_M^{-1} M\right)^n}{n}(-1)^{n+1}\right]. 
\label{VTr}
\EE
yielding a massive expansion for the fermions.
The GEP contains only graphs up to first order and is given by the first two terms, the two fermion loops
in Fig.~\ref{fig1}. The first term in the expansion, the zeroth-order loop, is
\BE
V_\Psi^{(0)}=i \Tr\int \ppp \log ( \fsl p-M)=-4 I_1(M)
\label{VPsi0}
\EE
while the second term, the crossed first-order loop, by Eq.(\ref{dI1}) reads
\BE
V_\Psi^{(1)}=-M\frac{\partial}{\partial M} V_\Psi^{(0)}=4 M^2 I_0(M).
\label{VPsi1}
\EE
We observe that, without the crossed graph, the one-loop vacuum energy 
would be given by $V_\Psi^{(0)}=-4 I_1(M)$ which is unstable and unbounded from below according to Eq.(\ref{IDR}).
On the other hand, with one counterterm insertion, the first-order crossed graph makes the GEP
bounded and yields the total first order effective potential
\BE
V_\Psi=-4\left[I_1(M)-M^2 I_0(M)\right]
\label{VPsiG}
\EE
which is exactly the GEP found in Ref.~\citenum{stevenson86} by a direct variational method, provided that we take
the chiral limit and set the external gluon field to zero. By dimensional regularization, inserting Eq.(\ref{IDR}),
the quark contribution to the GEP reads
\BE
V_\Psi(M)=\frac{3 M^4}{16\pi^2}\left[\log\frac{M^2}{\Lambda^2} +\frac{1}{6}\right]
\label{VPsiL}
\EE
and has a minimum at $M_0^2=\Lambda^2 e^{-2/3}$ where $V_\Psi (M_0)=-3M_0^4/(32\pi^2)<0$.

\section{Discussion}\label{disc}

Let us summarize the main findings of the previous sections. 

The GEP for the GL model of superconductivity,
namely $U(1)$ scalar electrodynamics, is recovered by a more general analysis based on a massive expansion,
yielding a mass generation even when the original model is scaleless.
The derivation of the GEP for pure $SU(N)$ Yang-Mills theory and chiral QCD also gives an original independent
way to introduce the massive expansion: a change of the expansion point with massive propagators in the internal lines
of the loops. The expansion acquires an evident variational meaning and emerges from the same variational argument
that leads to the GEP. However, while the GEP is limited because of its first-order nature that leaves the fermions
decoupled, in the massive expansion higher order terms can be easily included, yielding a powerful analytical tool for
the study of QCD in the infrared and providing two-point functions that are in very good agreement with the results
of lattice simulations\cite{ptqcd,ptqcd2,analyt,scaling}.

That said, the GEP gives a variational proof for chiral symmetry breaking and dynamical mass generation. Even if
the actual values of the masses cannot be trusted because the quarks are decoupled, the variational nature of the
calculation gives a proof that the vacuum of massless gluons and quarks is not stable. The Yang-Mills effective
potential is given by the function $U(\alpha,x)$ of Eq.(\ref{U}) and is shown in  Fig.~\ref{fig3}. An interesting
feature is the occurrence of an unstable relative minimum at $m=0$ and a stable minimum at $m>0$. We could 
speculate and see an analogy with the double solution that occurs in the Dyson-Schwinger formalism: an unphysical
massless scaling solution and a physical massive gluon propagator. Even if decoupled, the quark term of the GEP has
an absolute minimum at a finite $M>0$ according to Eq.(\ref{VPsiL}), predicting the breaking of chiral symmetry of
QCD.

We can see the absolute minimum of the GEP
as a best expansion point for the massive expansion. In that sense, it is relevant to note that, once the crossed
graph is included, the quark term of the GEP is also bounded from below. In fact, the counterterm keeps trace of
the scaleless nature of the original Lagrangian and is needed for imposing that the Lagrangian is not
modified in the expansion. 

We are left with two independent mass parameters, $m$ and $M$, that must be determined by
the phenomenology since their explicit expressions depend on the unknown renormalized scale $\Lambda$.
Assuming that the scale $\Lambda$ is the same in the gluon and quark sector, which is not obvious,
the minimum of the GEP would give a best ratio of masses by Eqs.(\ref{GEPYM}),(\ref{VPsiL}) 
\BE
\frac{M_0}{m_0}=e^{\>\left( \displaystyle{ \frac{1}{3}+\frac{2}{3\alpha}} \right) }
\label{ratio}
\EE
linking together the dynamical generation of the gluon mass with the chiral symmetry breaking.
While highly non-perturbative and non analytic in the limit $\alpha\to 0$,
the suggested ratio of Eq.(\ref{ratio}) suffers the limitations of the quark-gluon
decoupling in the GEP and can be only regarded as a  starting point for  more refined
calculations.

\bibliographystyle{ws-procs9x6} 


\end{document}